\documentclass{aa}
\input psfig.sty
\begin{document}
\title{The Radio SNR G65.1+0.6 and its associated pulsar J1957+2831}
\author{W.W. Tian \inst{1,2}
\and
      D.A. Leahy \inst{2}}
\authorrunning{Tian $\&$ Leahy}
\offprints{Wenwu Tian}
\institute{National Astronomical Observatories, CAS, Beijing 100012, China \\
\and
Department of Physics \& Astronomy, University of Calgary, Calgary, Alberta T2N 1N4, Canada}
 
\date{Received XX 2006; accepted xx, 2006} 
\abstract{New images of the radio Supernova Remnant (SNR) G65.1+0.6 are presented, based on the 408 MHz and 1420 MHz continuum emission and the HI-line emission data of the Canadian Galactic Plane Survey (CGPS). A large shell-like structure seen in the 2695 MHz Effelsberg map appears to have nonthermal spectral index. HI observations show 
structures associated with the SNR G65.1+0.6 in the radial velocity range of -20 to -26 km$/$s and suggest a distance of 9.2 kpc for the SNR.  The estimated Sedov age for G65.1+0.6 is 4 - 14$\times$10$^{4}$yr. The pulsar (PSR) J1957+2831 is possibly associated with G65.1+0.6, with consistent distance and kinematic age estimate, but different characteristic age than the SNR. The EGRET source 3EG J1958+2909 and $\gamma$-ray source 2CG 065+00 are also near the eastern edge of the SNR but do not agree in position with the pulsar and are likely not associated with the SNR.  The SNR's flux densities at 408 MHz (8.6$\pm$0.8 Jy), 1420 MHz (4.9$\pm$0.5 Jy) and 2695 MHz (3.3$\pm$0.5 Jy) have been corrected for flux densities from compact sources within the SNR. The integrated flux density based spectral index (S$_{\nu}$$\propto$$\nu$$^{-\alpha}$) between 1420 MHz and 408 MHz is  0.45$\pm$0.11 and agrees with the T-T plot spectral index of 0.34 $\pm$0.20. The nearby SNR DA495 has a T-T plot spectral index of 0.50$\pm$0.01.    

}
\titlerunning{The SNR G65.1+0.6 and its associated pulsar J1957+2831}
\maketitle 

\section{Introduction}
Since G65.1+0.6 was discovered as a low surface brightness SNR (Landecker et al., 1990), little research on this SNR has been done. Due to lack of independent distance and age estimates for G65.1+0.6, it was difficult for previous authors to compare the SNR with nearby discovered pulsars (Lorimer et al., 1998). The detections of an EGRET source 3EG J1958+2909 (Hartman et al., 1999) and a TeV $\gamma$-ray source 2CG 065+00 (Alexandreas et al., 1991; Swanenburg et al., 1981) which partly overlap the SNR G65.1+0.6 give increased reason to further study G65.1+0.6.  In this paper, we present the SNR's continuum images at higher sensitivity and resolution than previously at 408 MHz and 1420 MHz, and first investigate HI-line emission at various radial velocities for detecting interactions of the remnant with the surrounding gas, estimate its distance and age, and access its possible association with the nearby EGRET source, $\gamma$-ray source and pulsar. 
We see evidence for a large SNR-like shell in the 2695 MHz Effelsberg map, and investigate the evidence below.
\section{Observations and Analysis}

The continuum and HI emission data sets come from the CGPS,
which is described in detail by Taylor et al. (2003).
The data sets are mainly based on observations from the Synthesis Telescope 
(ST) of the Dominion Radio Astrophysical Observatory (DRAO). The spatial
resolution of the continuum images of G65.1+0.6 is 0.8$^{\prime}$ $\times$ 1.7$^{\prime}$ at 1420 MHz and 2.8$^{\prime}$$\times$5.9$^{\prime}$ at 408 MHz. The synthesized beam for 
the HI line images is 58$^{\prime}$$^{\prime}$$\times$ 2.0 $^{\prime}$ and the radial velocity 
resolution is 1.32 km$/$s. DRAO ST observations 
are not sensitive to structures larger than an angular 
size scale of about 3.3$^{o}$ at 408 MHz and 56$^{\prime}$  at 1420 MHz. Thus the CGPS includes 
data from the 408 MHz all-sky survey of Haslam et al (1982), sensitive to structure greater than 51$^{\prime}$, and the Effelsberg 1.4 GHz Galactic plane survey 
of Reich et al. (1990, 1997), sensitive to structure with resolution 9.4$^{\prime}$ for large scale emission 
(the single-dish data are freely available by http://www.mpifr-bonn.mpg.de/survey.html). 
The low-order spacing HI data is from the single-antenna survey of the 
CGPS area (Higgs $\&$ Tapping 2000) with resolution of 36$^{\prime}$.  See Taylor et al. (2003) for detail of the method of combining the synthesis telescopes and single dish observations. 

We analyze the continuum and HI images of G65.1+0.6 and determine its flux
densities and distance.  
 For G65.1+0.6, integrated flux density's errors are found by comparing results for several different choices of background region. For compact sources, the flux density's errors are taken as the formal Gaussian fit errors.  
The influence of compact sources within the SNR is much reduced by employing similar methods to Tian and Leahy (2005).
 
\section{Results}
\begin{figure*}
\vspace{45mm}
\begin{picture}(200,300)
\put(-110,500){\includegraphics{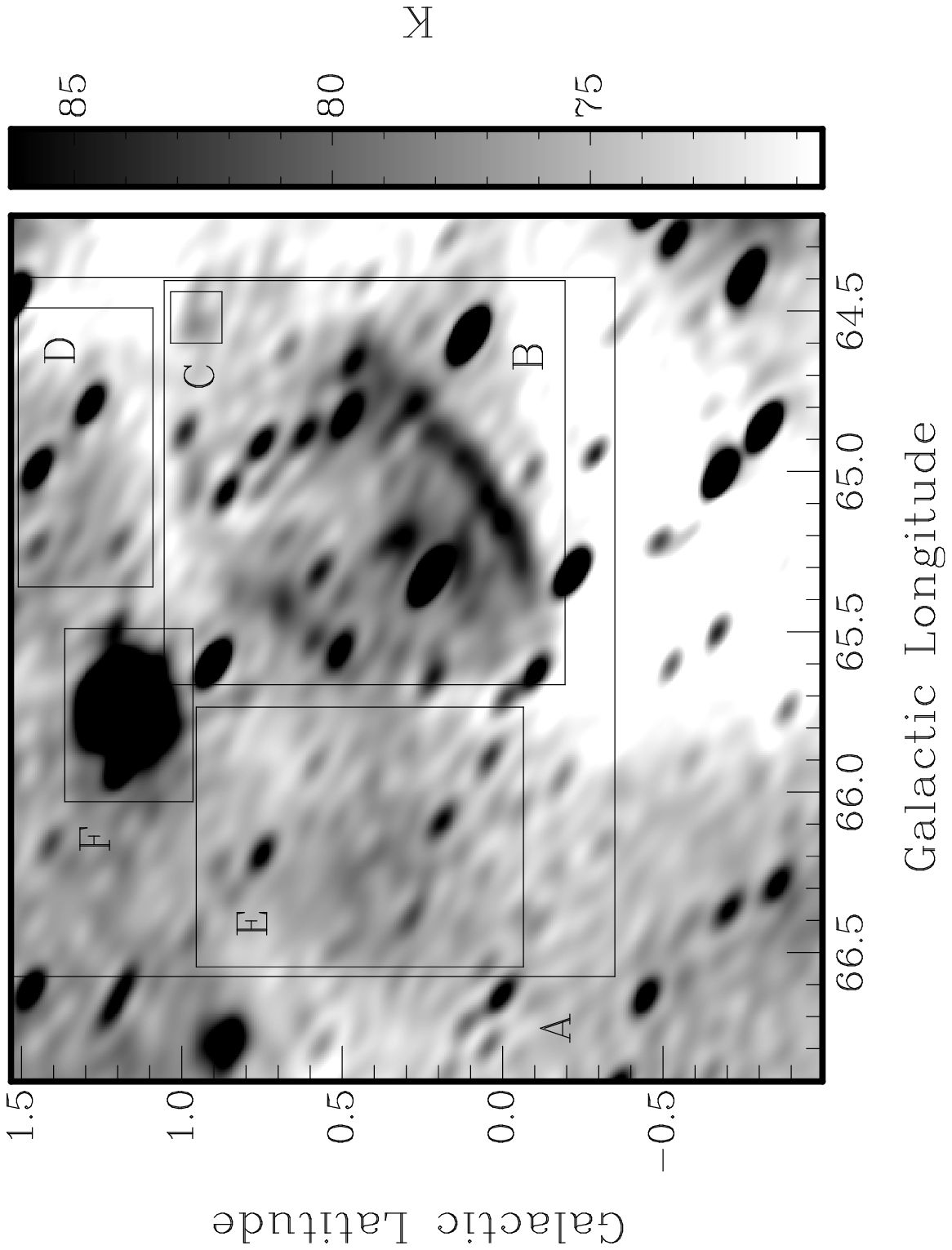}}
\put(211,90){\includegraphics{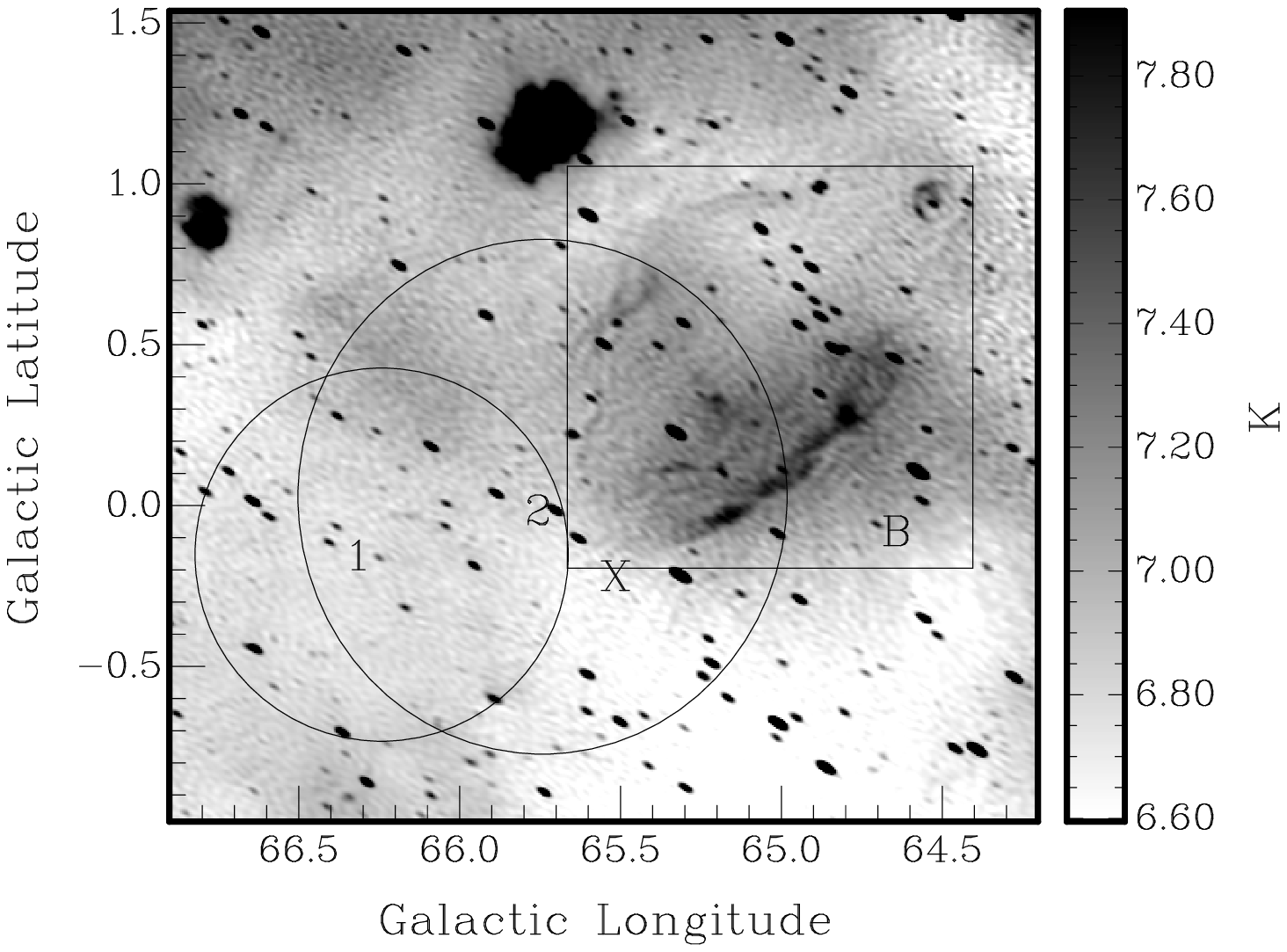}}
\put(-110,280){\includegraphics{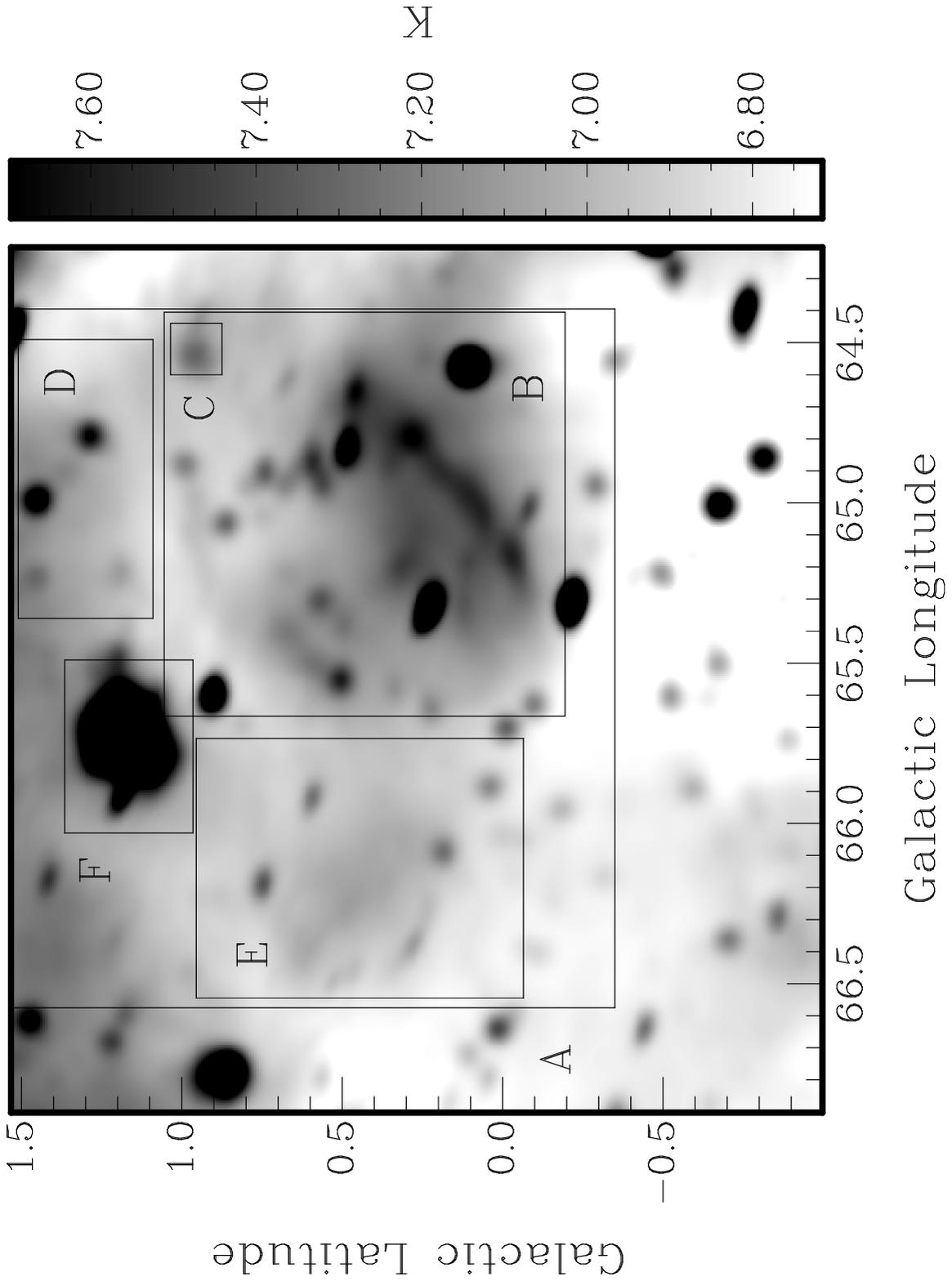}}
\put(260,-65){\includegraphics{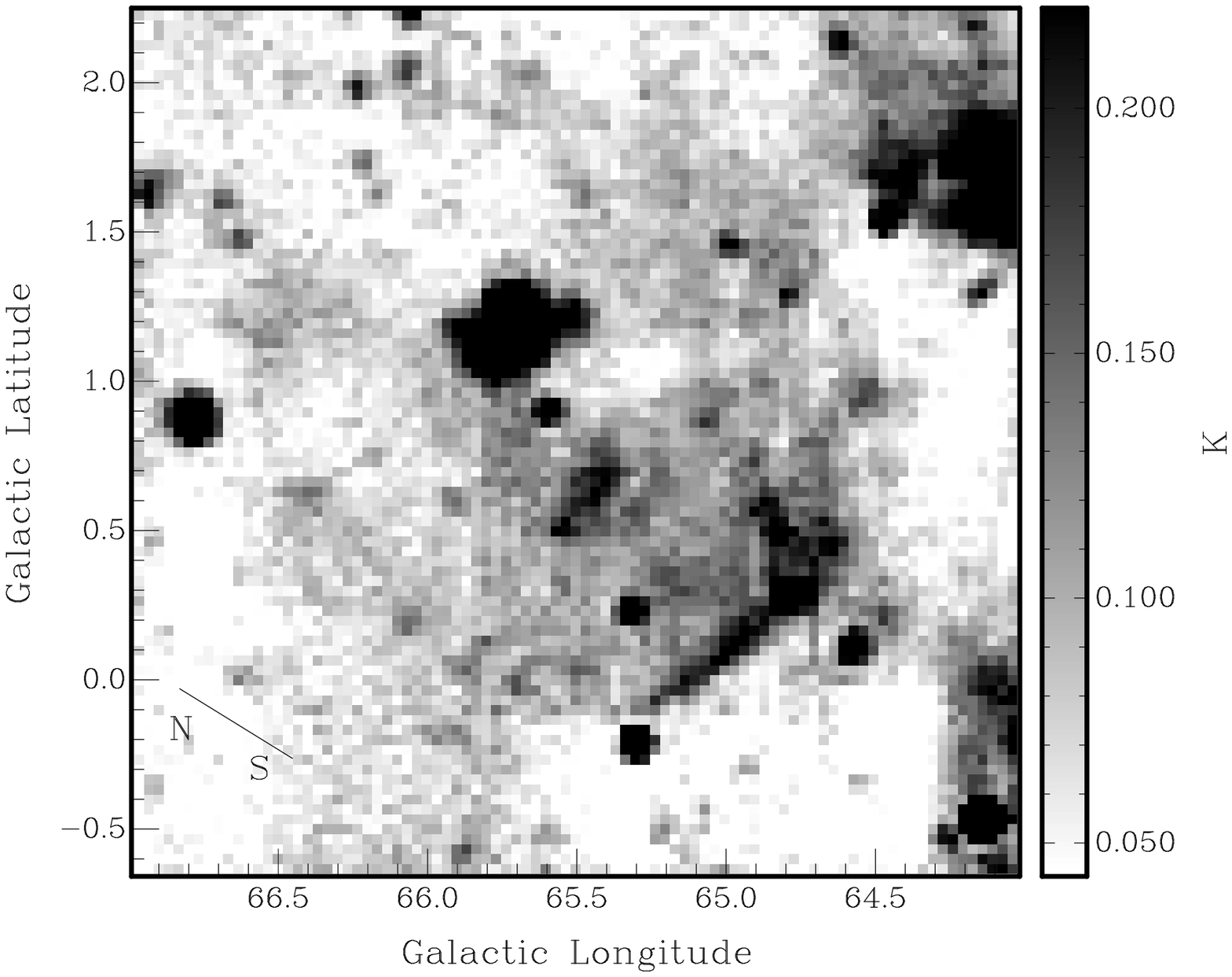}}
\end{picture}
\caption[xx]{The first row of images shows the CGPS maps at 
408 MHz (left) and 1420 MHz (right). The left of the second row shows the 1420 MHz map convolved to the same resolution as the 408 MHz map. 
The right of the second row is the 2695 MHz Effelsberg map.  
The 6 boxes labeled with letters A-F and used for T-T plots are shown in the left. The right of the first row shows positions and error circles of EGRET source 3EG J1958+2909 (95$\%$ confidence level, marked by number 1) and $\gamma$-ray 2CG 065+00 (90$\%$ confidence level, marked by number 2). Pulsar J1957+2831 is marked by letter X.
The direction of North (N) and South (S) is marked on the lower right image.}
\end{figure*}

\begin{figure}
\vspace{55mm}
\begin{picture}(0,50)
\put(-70,250){\includegraphics{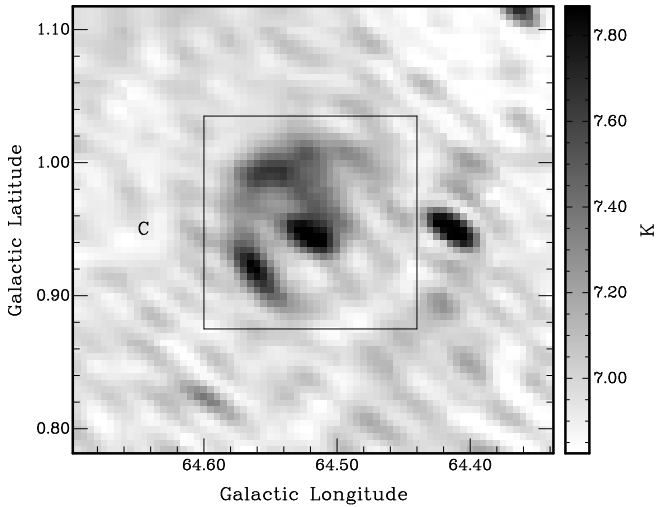}}
\end{picture}
\caption[xx]{Image of box C at 1420 MHz}
\end{figure}

\subsection{Continuum Emission}

The CGPS continuum images at 408 MHz and 1420 MHz are shown in the upper left and right panels of Fig. 1. The 6 boxes labeled with letters A-F and used for T-T plots are shown in the left. The upper right of Fig. 1 shows positions and error circles of EGRET source 3EG J1958+2909 (marked by number 1) and $\gamma$-ray 2CG 065+00 (marked by number 2). Pulsar J1957+2831 is marked by letter X. 
The lower left shows the 1420 MHz map convolved to the same resolution as the 408 MHz map.
The 2695 MHz Effelsberg map is reproduced in the lower right for reference
(F\"urst et al., 1990). The Effelsberg map has a resolution of 4.3$^{\prime}$ and a sensitivity of 50 mKT$_{B}$.

The 408 MHz image for G65.1+0.6 reveals more structures in comparison with Landecker's (1990) 408 MHz image, especially, side-lobe effects which are obvious in Landecker's maps are much reduced on our image. 
The 1420 MHz map shows much better detail of the fine structure in 
G65.1+0.6 than any previous image. More compact sources within the SNR are resolved in the 1420 MHz map.  More filament structures appear in the northern, northwestern and southwestern parts of the SNR.  Fig. 2 shows that a ring-shape structure  inside box C  is clearly detected near the western diffuse emission regions of the SNR.  The outlines of G65.1+0.6 at 408 MHz and 1420 MHz are very similar. 

The 3$^{0}$$\times$3$^{0}$ 2695 MHz Effelsberg map appears to show a  circular-shell (Radius about 1$^{0}$) SNR-like emission region including the two known SNRs G65.1+0.6 and G65.7+1.2 (DA495). But the CGPS maps at both 408 MHz and 1420 MHz don't show this circular shell.  We mark the large emission region (A) and its two inner areas (D and E) related with G65.1+0.6 in order to study the emission further.   

\begin{table*}
\begin{center}
\caption{List of the 16 brightest compact sources ( the first 11 sources are within G65.1+0.6) and their integrated flux densities within box A}
\setlength{\tabcolsep}{1mm}
\begin{tabular}{cccccccc}
\hline \hline
 No.&RA(2000) & Dec(2000)& GLON&GLAT & S$_{408MHz}$&S$_{1420MHz}$& Sp. Index\\
\hline & [h m s]&[$^{o}$' "]&deg &deg &mJy&mJy&$\alpha$\\
\hline
\hline 
1& 19 53 46.53& 27 52 56.6& 64.566&0.109 &1388 $\pm$42&533$\pm$ 17&0.77 (0.73 to 0.80)\\
2& 19 55  3.77& 28 35 43.2& 65.323&0.234 &1044 $\pm$34&318$\pm$ 10&0.95 (0.92 to 0.99)\\
3& 19 53  3.10& 29 10 30.2& 65.593&0.910 & 610 $\pm$19&261$\pm$  9&0.68 (0.65 to 0.72)\\
4& 19 52 53.01& 28 17 43.0& 64.819&0.490 & 267 $\pm$20&111$\pm$  4&0.71 (0.64 to 0.78)\\
5& 19 57  2.74& 28 39 52.6& 65.608&-0.102& 229 $\pm$16& 37$\pm$  3&1.45 (1.36 to 1.55)\\
6& 19 52  3.14& 28 29 46.9& 64.898&0.750 & 191 $\pm$37& 31$\pm$  7&1.46 (1.12 to 1.80)\\
7& 19 56 11.84& 27 59 46.5& 64.940&-0.290& 147 $\pm$ 9& 61$\pm$  6&0.70 (0.61 to 0.81)\\
8& 19 51 57.62& 28 41 48.9& 65.060&0.870 & 137 $\pm$30& 44$\pm$  3&0.91 (0.66 to 1.13)\\
9& 19 54 31.32& 28 55 59.3& 65.551&0.510 & 127 $\pm$10& 49$\pm$  3&0.76 (0.69 to 0.84)\\
10& 19 57 7.54& 28 58 25.4& 65.881&0.044 &  94 $\pm$10& 46$\pm$  2&0.58 (0.49 to 0.67)\\
11& 19 56 51.73& 28 46 24.5& 65.680&-0.011&  83 $\pm$13& 60$\pm$  4&0.26 (0.13 to 0.41)\\
\hline
12& 19 56 46.80& 28 21 33.9& 65.317&-0.211& 765 $\pm$25&577$\pm$340&0.23(-0.15 to 0.94)\\
13& 19 49 27.70& 28 55 57.2& 64.984&1.461 & 290 $\pm$12&135$\pm$  7&0.62 (0.57 to 0.67)\\
14& 19 49 39.44& 28 40 27.0& 64.783&1.293 & 281 $\pm$12&100$\pm$  4&0.83 (0.78 to 0.87)\\
15& 19 55  3.55& 29 36  6.0& 66.184&0.755 & 117 $\pm$ 6& 56$\pm$  2&0.75 (0.70 to 0.80)\\
16& 19 52 22.96& 29 56  1.9& 66.170&1.424 &  48 $\pm$ 6& 42$\pm$  1&0.11 (0.01 to 0.21)\\
\hline
\hline
\end{tabular}
\end{center}
\end{table*}

\begin{table}
\begin{center}
\caption{408-1420 MHz T-T plot spectral indices
with and without Compact Sources(CS)}
\setlength{\tabcolsep}{1mm}
\begin{tabular}{ccc}
\hline
\hline
 Sp. Index   |    &$\alpha$ &  $\alpha$  \\
\hline
\hline
 Area \vline &including CS &  CS removed \\
\hline
 A& $0.30\pm$0.21& 0.25$\pm0.22$\\
 B& $0.43\pm$0.21& 0.34$\pm0.20$\\
 C& $0.35\pm$0.13& 0.35$\pm0.13$\\
 D& 0.59$\pm$0.13& 0.51$\pm$0.20\\
 E& 0.34$\pm$0.59& 0.18$\pm$1.20*\\
 F& 0.50$\pm$0.01&0.50$\pm$0.01\\
\hline
 \hline
\end{tabular}
\\**Manual fitting gives a more reliable value 0.7.
\end{center}
\end{table}

\begin{table}
\begin{center}
\caption{Integrated flux densities and 408-1420 MHz spectral indices of G65.1+0.6 and compact sources within the SNR}
\setlength{\tabcolsep}{1mm}
\begin{tabular}{cccc}
\hline
\hline
Freq.& G65.1+0.6& CS of G65.1+0.6 &G65.1+0.6 and CS\\
\hline
MHz &Jy&Jy &Jy\\
\hline
\hline
 408& 8.6$\pm$0.8&4.3$\pm$0.2&12.9$\pm$1.0\\
 1420&4.9$\pm$0.5&1.6$\pm$0.1& 6.5$\pm$0.6\\
 2695&3.3$\pm$0.5&0.9$\pm$0.1& 4.2$\pm$0.4\\ 
\hline
$\alpha$&0.45$\pm$0.11&0.79$\pm$0.06&0.55$\pm$0.10\\
\hline
\hline
\end{tabular}
\end{center}
\end{table}

\subsection{T-T Plot Spectral Indices}
Bright compact sources affect the measured integrated flux densities for G65.1+0.6 and its measured spectral index. Thus we correct for the effects of compact sources.  Table 1 lists properties of the 16 brightest compact sources which are detected in box A, and the first 11 of which are 
within G65.1+0.6 at both 408 MHz and 1420 MHz.   

First we discuss spectral indices between 408 MHz and 1420 MHz based on 
the T-T plot method (Turtle et al., 1962).
The principle of the T-T plot method is that spectral indices 
(T$_{\nu}$=T$_{o}$$\nu$$^{-\beta}$) are calculated from a fit of a linear 
relation to the T$_{1}$-T$_{2}$ values of all pixels within a given map region. 
T$_{1}$ is the brightness temperature of a map pixel at one frequency and 
T$_{2}$ is for the second frequency. The higher resolution image has been smoothed to the lower resolution for the T-T plot comparison. The brightness temperature spectral 
index $\beta$ is derived from the slope of the line. The error in spectral index is derived from the uncertainly in slope of the line. 
The flux density spectral index $\alpha$ 
(S$_{\nu}$$\propto$$\nu$$^{-\alpha}$) 
is related to $\beta$ by $\beta$=$\alpha$+2. Spectral index 
refers to flux density spectral index $\alpha$ in this paper unless specifically noted otherwise. 

For the T-T plot spectral index analysis, we select a single large region (A) including the two SNRs G65.1+0.6 and DA495, region B including G65.1+0.6, region F including DA495 alone and other 3 areas (C, D, E) around G65.1+0.6, as shown in Fig. 1. Region A yields the T-T plots shown in the upper half of Fig. 3. Region B yields the T-T plots shown in the lower half of Fig. 3.

In the T-T plot analysis, two cases are considered: using all pixels including compact sources; and excluding compact sources listed in Table 1 from the images. 
The compact sources are bright compared to the SNR emission. 
Since the compact sources have a steeper spectrum than the SNR, 
they are seen in the T-T plot (Fig. 3 left panel) as the steeper lines of 
points extending to higher $T_B$. 
We completely remove regions of pixels including 
the compact sources from the analysis. Each region is taken to be a few 
beamwidths across, so that any contribution from the compact source is below 
1$\%$ of the diffuse SNR emission.

\begin{figure*}
\vspace{110mm}
\begin{picture}(60,140)
\put(-42,450){\includegraphics{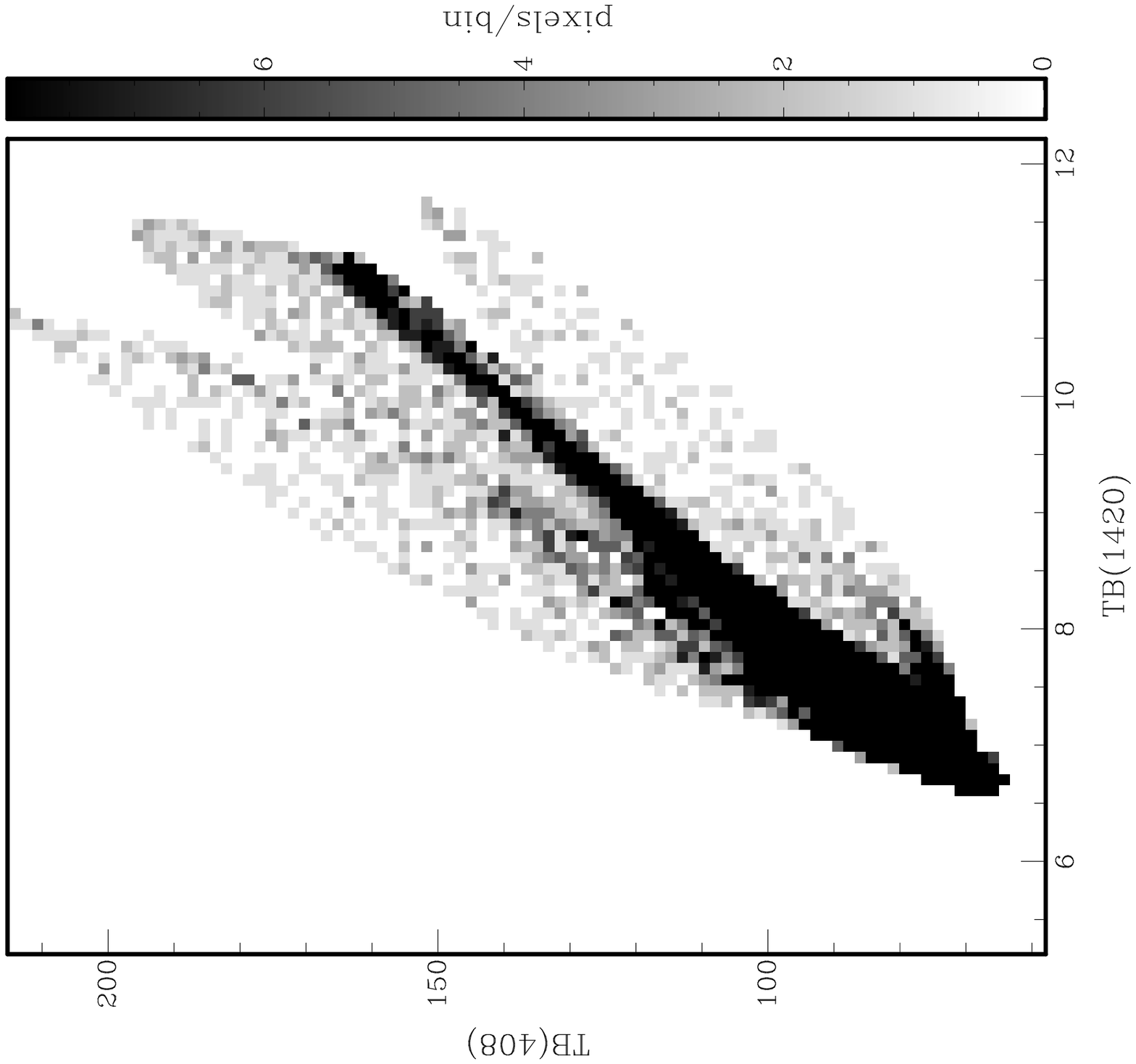}}
\put(210,452){\includegraphics{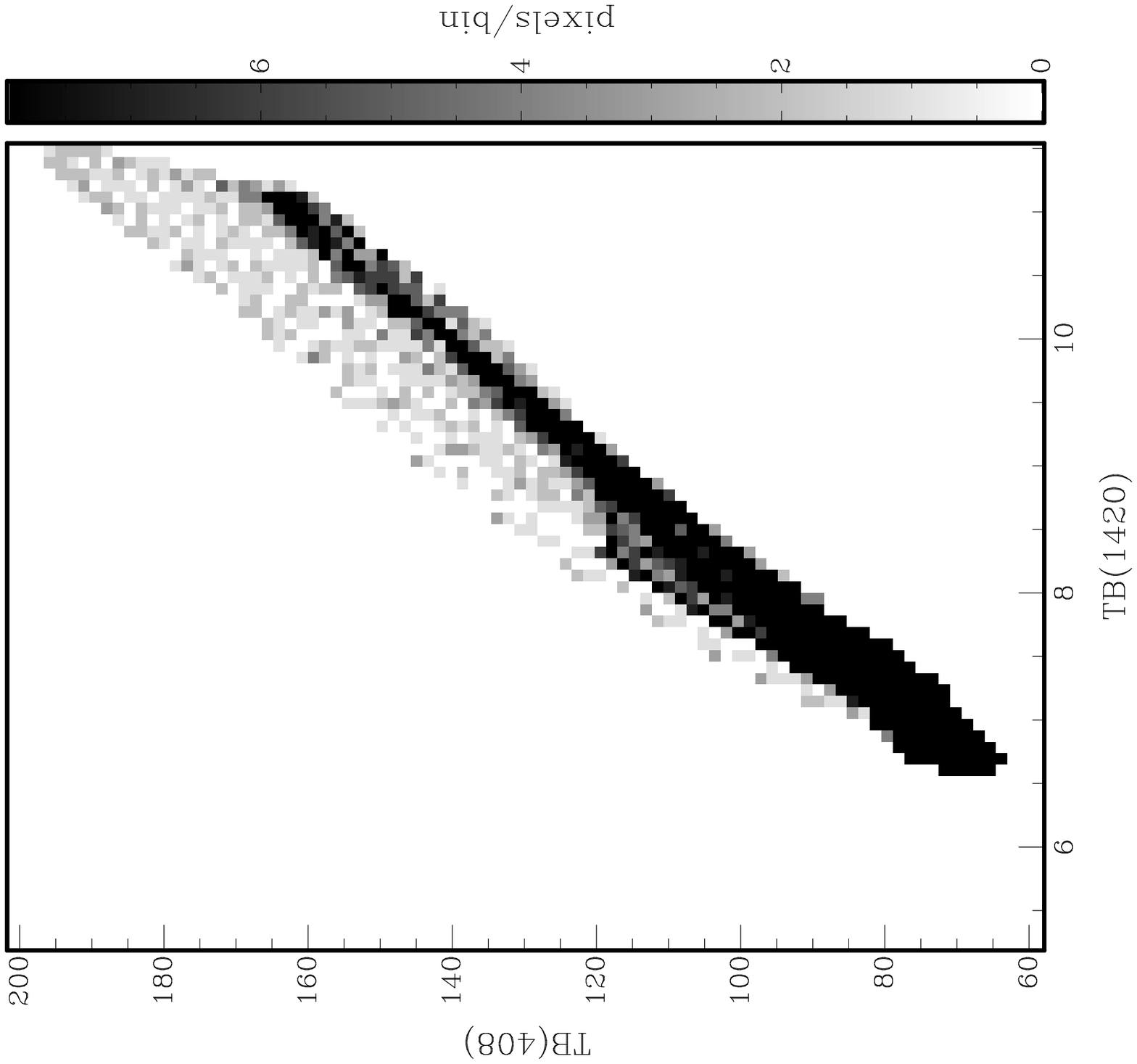}} 
\put(-50,230){\includegraphics{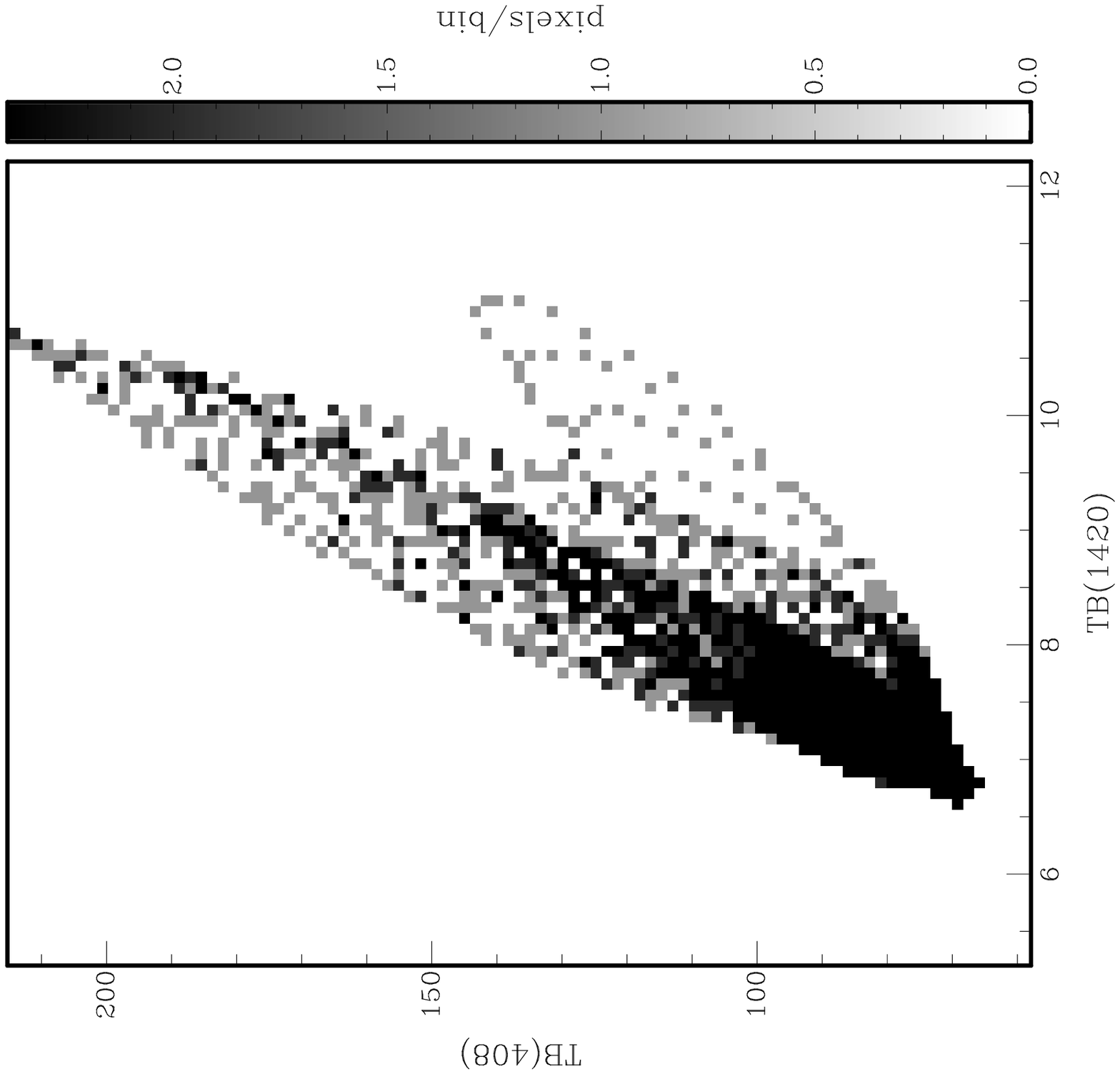}} 
\put(210,232){\includegraphics{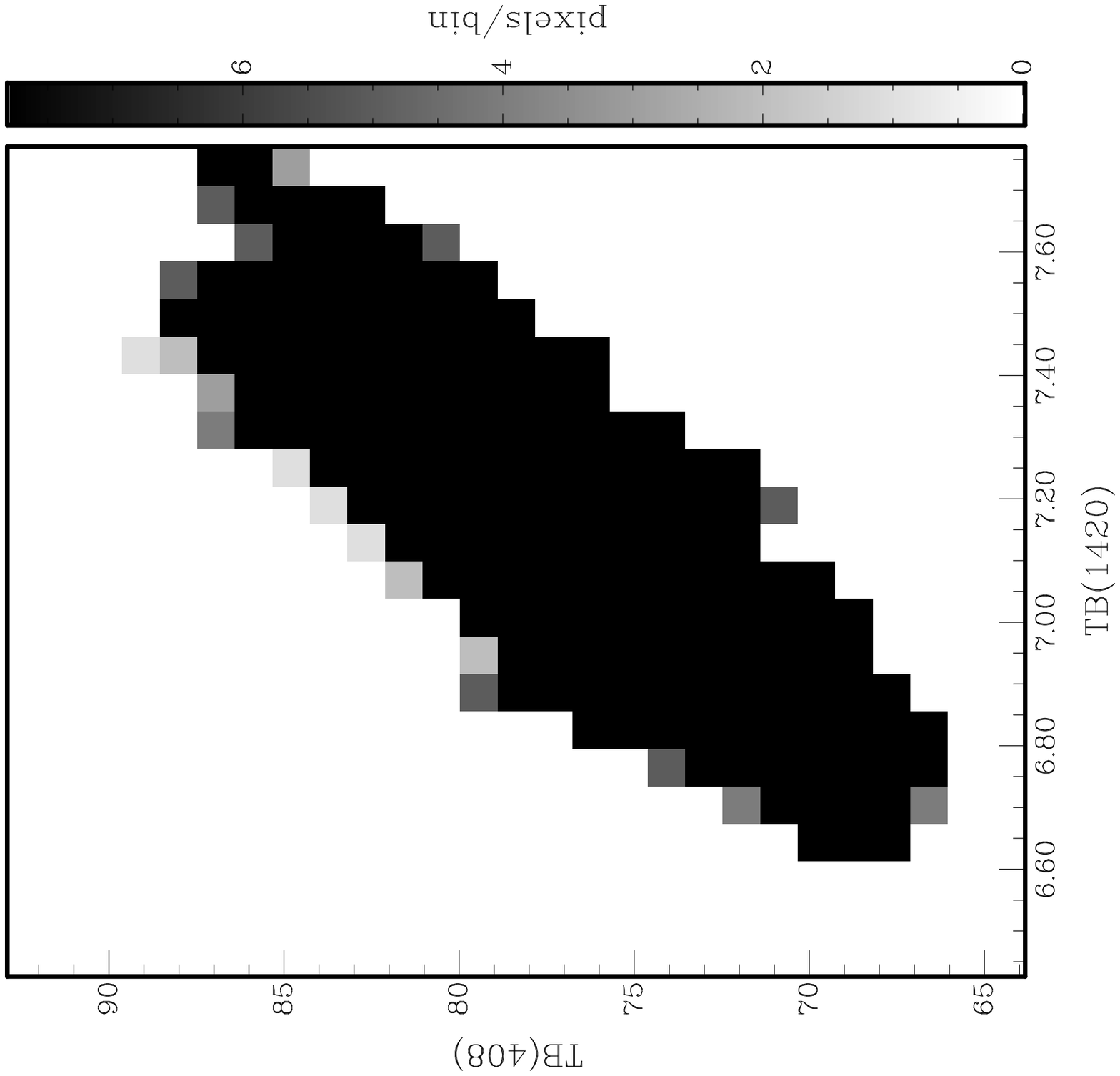}}
\end{picture}
\caption[xx]{408-1420 MHz T-T plots for region A (the upper half) and region B (the lower half). The left plots for maps including compact sources; The right plots for compact sources removed from analysis}
\end{figure*}

Table 2 lists the results for two cases of analysis: 
including compact sources and removing compact sources. 
Visual inspection of the T-T plots confirms that the second method produces the most reliable results.  The compact sources' influence on the spectral index calculation
is obvious in the T-T plots of Fig. 3, and also seen in Table 2 for areas A, B, D and E.
From now on we discuss spectral indices derived with compact sources removed, unless specified otherwise.

Region A includes several emission regions. The spectral index for a box F including DA495 alone is given in Table 2 (0.50$\pm$0.01). From the right panels of Fig. 3, it is obvious that the T-T plot of G65.1+0.6 is only left lower part of region A's T-T plot. The upper right part of region A's plot is from the SNR DA495.

\begin{figure}
\vspace{50mm}
\begin{picture}(0,100)
\put(-20,-25){\includegraphics{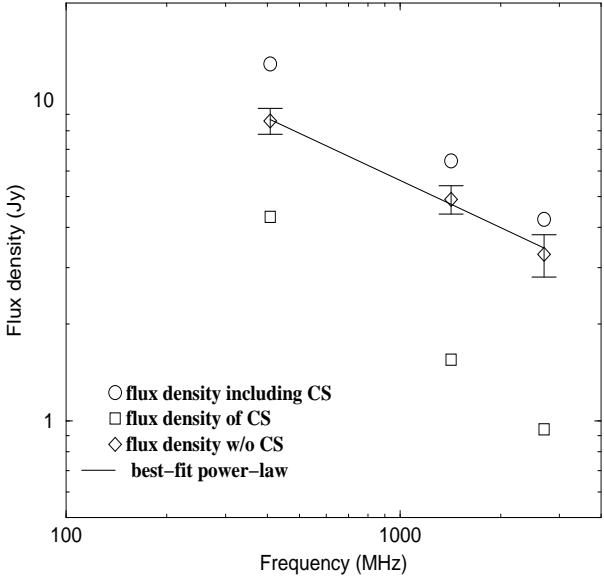}}
\end{picture}
\caption[xx]{Radio spectrum of G65.1+0.6 with and without Compact Sources (CS). The power law model has a best-fit spectral index of 0.49 with 1$\sigma$ uncertainty range of 0.40 to 0.57.}
\end{figure}

\begin{figure*}
\vspace{120mm}
\begin{picture}(60,60)
\put(-10,-65){\includegraphics{hi-20.eps}} 
\put(112,-65){\includegraphics{hi-21.eps}} 
\put(233,-65){\includegraphics{hi-22.eps}} 
\put(355,-65){\includegraphics{hi-22-1.eps}}
\put(-10,64){\includegraphics{hi-23.eps}} 
\put(112,64){\includegraphics{hi-24.eps}} 
\put(232,64){\includegraphics{hi-25.eps}}
\put(356,64){\includegraphics{hi-26.eps}}
\put(-40,444){\includegraphics{hi-41.eps}}
\put(82,444){\includegraphics{hi-42.eps}}
\put(202,444){\includegraphics{hi-45.eps}}
\put(326,444){\includegraphics{hi-46.eps}}
\end{picture}
\caption[xx]{HI emission in the field centered on G65.1+0.6 from -20 to -26 km$/$s and -41 to -46 km$/$s. The radial velocity of each map is indicated at its top left corner. The outline of G65.1+0.6 in 1420 MHz continuum emission
is indicated by the contour at 7.2 KT$_{B}$.}
\end{figure*}

\subsection{Integrated Flux Densities and Spectral Indices}

We have derived integrated flux densities of G65.1+0.6 from the 
408 MHz and 1420 MHz maps.  Values given have diffuse background subtracted.
The resulting 408 MHz to 1420 MHz spectral index, using flux densities without compact sources, is 0.45$\pm$0.11.  
Table 3 lists the flux densities and spectral indices of G65.1+0.6 and 
the compact sources within G65.1+0.6. 
Compact sources contribute about 33$\%$ at both 408 MHz and 25$\%$ at 1420 MHz to the SNR's flux densities, and have a significant effect on the spectral index. 
It is noted that the SNR spectral index derived from integrated flux densities is consistent with the SNR spectral index (0.34$\pm$0.20) derived by the T-T plot method.

We have calculated total compact source flux densities for 2695 MHz 
frequencies, using the 408-1420 MHz spectral index upper and lower limits and
flux densities from Table 1.  The compact sources contribute 21$\%$ at 2695 MHz to the SNR's flux densities.  
We have recalculated the flux density values of G65.1+0.6 at 2695 MHz by subtracting the compact source flux density. 
 We note that F\"urst et al (1984) gave a flux density based 
on the 2965 MHz Effelsberg image with resolution 4.4 $\times$ 4.4 arcmin, but we obtain a new value from the Effelsberg 2695 MHz image  
with a little higher resolution 4.3 $\times$ 4.3 arcmin: 3.3$\pm$0.5 Jy and use this instead.
We fit the resulting flux density values with a power-law to obtain spectral index.  Figure 4 shows the corrected flux densities and the best-fit power-law.  
The best fit spectral index is 0.49 with 1$\sigma$ uncertainty range
of 0.40 to 0.57.  The fit to G65.1+0.6 spectrum including compact sources flux density gives $\alpha$=0.58. The steeper $\alpha$ is expected since  compact source has the steeper spectrum than SNR (see Table 3).

\subsection{HI Emission}
We have searched the CGPS radial velocity range for features in the HI which might 
relate to the morphology of G65.1+0.6.  There is emission which is coincident with the boundary of G65.1+0.6 in the velocity
range -20 to -26 km/s, and only in this velocity range.
 The spatial relation with the
edge of G65.1+0.6 indicates that the HI is associated with
the SNR.  These HI observations show a good association of HI 
features with G65.1+0.6, similar to other accepted HI associations with 
SNRs and with similar HI velocity ranges ($\sim$10km/s), e.g.  HI associated with G126.2+1.6 (Tian $\&$ Leahy, 2006), with G127.1+0.6 (Leahy $\&$ Tian, 2006) and with DA530 (Landecker et al., 1999).  
Fig. 5 shows maps of HI emission in twelve channels. Each map has superimposed on it contours of continuum emission at 1420 MHz 
chosen to show G65.1+0.6. 

\subsection{EGRET Source, $\gamma$-Ray and Pulsar nearby the SNR}
The upper right of Fig. 1 shows positions of the EGRET source 3EG J1958+2909 detected by Hartman et al.( 1999), the TeV $\gamma$-ray source 2CG 065+00 detected by Alexandreas et al.(1991) and the PSR J1957+2831 discovered by Lorimer (1998). The EGRET and $\gamma$-ray sources overlap each other and the SNR G65.1+0.6. The EGRET and TeV sources are likely the same source with position inconsistent with the pulsar's position and are likely not associated with the SNR. The pulsar is nearby at the eastern boundary of the SNR G65.1+0.6. 

\section{Discussion}
\subsection{The Distance and Age to G65.1+0.6}
Since the velocity is negative (-20 to -26 km/s) this implies it is outside the solar circle and at l=65.1$^{0}$, is very far away.
We use circular galactic rotation velocity V$_{R}$=V$_{0}$=220 km/s. Then R$_{0}$/R=0.89 for -22 km/s, 0.87 for -26km/s and 0.90 for -20km/s.
Then we use R$_{0}$=8.5kpc, so d=9.2kpc (-22 km/s), 9.0kpc (-20km/s) and 9.6kpc (-26km/s).  Non-circular motions in the HI feature increase the distance uncertainty, eg $\pm$5 km/s,  increases the distance range to 8.7 - 10.1 kpc. This means G65.1+0.6 is very large (angular diameters 90$^{\prime}$ by 51$^{\prime}$, average diameter 70$^{\prime}$): yields radius 93.5pc (d=9.2kpc).

The SNR is of low surface brightness  and large radius, so probably exploded in a low density part of the ISM.
A Sedov model (Cox D., 1972) gives t=14$\times$10$^{4}$yr if n$_{0}$=0.01cm$^{-3}$, or
t=4.4$\times$10$^{4}$yr if n$_{0}$=0.001cm$^{-3}$. The radius at which cooling affects the dynamics of the SNR is 68pc if n$_{0}$=0.01cm$^{-3}$, or 127pc if n$_{0}$=0.001cm$^{-3}$. So it is possible that the SNR is still in the Sedov phase or has already entered the cooling phase depending on the density n$_{0}$. The shock temperature is 8$\times$10$^{5}$K if n$_{0}$=0.01 and 8$\times$10$^{6}$K if n$_{0}$=0.001. Since the density is low it would be faint in X-rays, and since it is quite distant it would be absorbed in X-rays by the intervening ISM.  The non-detection in the Rosat All-sky Survey of G65.1+0.6 is consistent with a large distance to G65.1+0.6.  

\subsection{Possible Association of the SNR G65.1+0.6 and the Pulsar J1957+2831}
 Lorimer et al. (1998) discovered PSR J1957+2831 which is located at the eastern edge of G65.1+0.6, and commented that the pulsar is not likely to be associated with G65.1+0.6 due to the large dispersion measure (DM) distance (7.0$\pm$2.3 kpc) and old characteristic age (1.6 Myr) of the pulsar, and their statistical grounds of pulsar-SNR pairs in the case of absence of G65.1+0.6's distance and age parameters. But Lorimer et al. (1998) also admitted a possibility that the PSR J1957+2831 may indeed be younger than its characteristic age because the pulsar has a spectral index typical for other young pulsars. 
If we assume that the pulsar birth-place is at geometrical centre of G65.1+0.6, and moves away toward east of the SNR at typical velocity of 500 km/s (Lyne et al., 1994), the kinematic age of the pulsar approximately is 18$\times$10$^{4}$yr and consistent with the estimated SNR's above age. Although previous studies (Tian $\&$ Leahy, 2004; Mereghetti et al., 2002; Kaspi et al., 2001) have suggested that pulsar characteristic ages can be poor age estimators for young pulsars, the pulsar J1957+2831 is not young. The pulsar's DM distance (7.0$\pm$2.3 kpc) and kinematic age agree with G65.1+0.6, and its characteristic disagrees with the age of the SNR, so it is possible that the pulsar is associated with the SNR.

\subsection{New Supernova Remnant G64.5+0.9?}
The 2695 MHz Effelsberg map appears to show a circular-shell SNR-like emission region most of which is included in region A shown in Fig. 1. The 408-1420 MHz T-T plot spectral index of areas C, D, E in Table 3 shows they are probably all nonthermal emission regions (each is about 2$\sigma$ different from thermal spectral index). This supports the possibility that there is a large diffuse SNR including the two known SNRs. But the HI images don't show any  association of HI features with the large circular shell. Thus current evidence is inconclusive. 

The ring-like structure in box C detected in the 1420 MHz image (Fig. 2) may be part of the western edge of G65.1+0.6. However, the ring has a good association with HI features in the velocity range -41 to -46 km/s as shown in Fig. 5. The HI velocity corresponds to a distance of 11kpc (-43 km$\/$s). Now we name the emission structure as a possible new SNR G64.5+0.9 which certainly needs to be confirmed by future high resolution and sensitivity observations. 

\section{Conclusion}  
Using the 408 MHz and 1420 MHz continuum emission and the HI-line emission data of the CGPS, physical parameters of G65.1+0.6: flux densities, spectral index, distance and age are either corrected or first obtained. Based on distance and age of both the PSR J1957+2831 and the SNR G65.1+0.6, we conclude that the pulsar and the SNR are possibly associated. Our analysis of G65.1+0.6 shows that the integrated flux density based spectral index between 1420 MHz and 408 MHz (0.45$\pm$0.11), T-T plot spectral index (0.34 $\pm$0.20) and the fit spectral index (0.49) are consistent. The nearby SNR DA495 has a T-T plot spectral index of 0.50$\pm$0.01.
A possible SNR G64.5+0.9 is suggested for its likely nonthermal emission feature, shell-type structure and associating with HI emission features. 
  
\begin{acknowledgements}
We acknowledge support from the Natural Sciences and Engineering Research Council of Canada. We thank Dr. R. Kothes at the DRAO provided technical help on the CGPS data processing. The DRAO is operated as a national facility by the National Research Council of Canada.  
The Canadian Galactic Plane Survey is a Canadian project with international partners. 
\end{acknowledgements}

\end{document}